 \documentclass[final,5p,times,twocolumn]{elsarticle}

\usepackage{amssymb}
\usepackage{lipsum}

\usepackage{amsmath}
\usepackage{graphicx}
\usepackage{epsf}
\usepackage{psfrag}
\usepackage{epsfig}
\usepackage{graphics}
\usepackage{amsfonts}
\usepackage{epstopdf}
\usepackage{slashed}
\usepackage{footnote}

\newcommand{\be}{\begin{equation}}
\newcommand{\ee}{\end{equation}}
\newcommand{\bq}{\begin{eqnarray}}
\newcommand{\eq}{\end{eqnarray}}
\newcommand{\ba}{\begin{align}}
\newcommand{\ea}{\end{align}}
\usepackage{slashed}
\usepackage{tabularx}
\usepackage{pifont}
\usepackage[mathscr]{euscript}

\usepackage[latin1]{inputenc}
\usepackage[english]{babel}
\usepackage[T1]{fontenc}
\usepackage{graphicx}
\usepackage{amsmath}
\usepackage{amsfonts,amssymb,ifthen,amsthm}
\usepackage{amsfonts}
\usepackage{amssymb}
\usepackage{subfigure}

\newcommand{\D}{{\cal D}}

\journal{Physics Letters B}

\begin{document}

\begin{frontmatter}



\title{Scale Anomaly in Supersymmetric Quantum Electrodynamics: Ward Identities, Supersymmetry Breaking, and Decoupling}


\author[first]{A. Yu. Petrov}
\affiliation[first]{organization={Universidade Federal da Paraíba}, Departamento de Física
            city={João Pessoa},
            postcode={58051-970}, 
            state={Paraíba},
            country={Brasil}}
            
\author[second]{A. R. Vieira}
\affiliation[second]{organization={Universidade Federal do Triângulo Mineiro}, Departamento de 
de Ciências Exatas e Educação
            city={Iturama},
            postcode={38280-000}, 
            state={Minas Gerais},
            country={Brasil}}            

\ead{alexandre.vieira@uftm.edu.br}

\begin{abstract}
We study the breaking of the scale symmetry both within classical and quantum frameworks in the supersymmetric quantum electrodynamics (SQED) by applying a four-dimensional regularization that preserves supersymmetry. At the classical level, scale symmetry is shown to be broken by mass terms. At the quantum level, it is demonstrated to be broken by a term proportional to the SQED beta function in the massless limit. It is also shown how this anomaly manifests itself in the low- and high-energy limits
and how the Appelquist-Carazzone theorem reproduces the usual quantum electrodynamics result in the limit of heavy selectrons.
\end{abstract}







\end{frontmatter}




\section{Introduction}
\label{s1}

Symmetries are treated as the most important cornerstones used to formulate field theories. Although they are usually considered at the classical level, they can be tested in precision experiments, like the Lorentz and CPT symmetry tests (see \cite{Kostelecky:2008ts} for a review), and checked beyond tree level. Originally, there was a common belief that if some symmetries of given field theory models take place at the classical level, they must continue to be valid at the quantum level. However, this statement is not always true. The breaking of classical symmetries under quantum corrections is called an anomaly. Some of 
them are measurable because they are related to physical processes, like the Adler-Bell-Jackiw (ABJ) anomaly  \cite{Bell, Adler}, related to the pion decay into two photons, and the trace anomaly, related to hadronic processes \cite{Ellis} and to the glueball spectrum \cite{Boschi}. The trace of the energy-momentum tensor, which is equal to the dilatation current conservation associated with scale symmetry,  acquires nontrivial corrections breaking this symmetry at the classical level by massive terms, while massless field theories are scale invariant. However, using these theories for short distances reveals that scale symmetry is also broken when loop corrections are taken into account \cite{Callan}.

The trace anomaly also manifests itself as a breaking of conformal symmetry in curved space-time when matter fields are coupled to a gravitational background \cite{Duff}. The computation performed with different regularization schemes reveals universal impacts of this symmetry breaking in the anomalous energy-momentum tensor trace \cite{Duff2}-\cite{IRegTA}. From the physical point of view, it is interesting to know whether the anomaly is an artifact of the regularization employed, caused by the fact that a regularization scheme can spuriously break symmetries of the theory, or if it is indeed a physical effect (for example, sometimes the dimensional regularization can generate Proca-like terms breaking gauge symmetry). Such a discussion certainly should be done, for example, in supersymmetric (SUSY) theories or within scenarios where the 
Lagrangian of the theory involves objects which are well defined only in an integer space-time dimension, like $\gamma_5$ matrices and Levi-Civita symbols, as occurs within perturbative generation of the Lorentz 
and CPT-violating operators of the Lorentz-violating (LV) extended QED, see e.g.  \cite{Colladay:1998fq}.  

In the usual quantum electrodynamics (QED), the anomalous breaking of the scale symmetry was found with the use of different regularization schemes \cite{Ellis, LoopReg, Porto}, and it is worth computing this anomaly in a SUSY version of QED. SUSY theories provide a particularly interesting framework for the study of scale anomalies. The energy-momentum tensor, the supersymmetry current, and the $R$-current can be embedded into supercurrent multiplets in such a way that all these three currents are related \cite{multiplet}. The Ferrara-Zumino construction shows how this relation is implemented \cite{Ferrara}. In particular, the relation between the energy-momentum trace, the $R$-current and the $\beta$-function leads to the {\it anomaly puzzle}. At the same time, computations of the abelian and non-abelian $\beta$-functions for SUSY theories demonstrated that the dependence on the regularization or renormalization scheme is relevant in the construction of the Novikov-Shifman-Vainstein-Zakharov (NSVZ) scheme. In ref. \cite{Puentes},  a non-perturbative analysis was performed for the computation of the $\beta$ function and the NSVZ structure, including mass- and momentum-dependent effects. At the level of SUSY Ward identities, it was shown in \cite{Bzowski, Katsianis} that the manifestation of quantum anomalies depends on the choice of the current multiplet and on the finite counterterms used to define the corresponding currents.

In SUSY theories, the breaking of scale or superconformal symmetries is related to the structure of the multiplets. In ref. \cite{Cordova}, it was shown that there is a non-trivial relation among conformal and 't Hooft anomalies, in such a way that apparently different anomalies can be organized in a common multiplet. While in ref. \cite{Batista}, authors show that only the superconformal symmetry is unambiguosly broken unlike gauge symmetry. In the presence of soft SUSY breaking, ref. \cite{Martin} stresses the importance of employing a regularization prescription compatible with SUSY.

Since anomalies are physical because they can be measured, some conventional regularization schemes, such as dimensional regularization, are not suitable for some studies in SQED, in particular, in diagrammatic anomaly computation. Actually, if these schemes are supplemented by special renormalization prescriptions, which restore the Slavnov-Taylor identities, their use is possible \cite{Slavnov}-\cite{Slavnov4}. Nevertheless, it is mandatory to find out if the symmetry breaking is spurious, {\it i.e.} caused by the regularization scheme, or physical indeed. Of course, the spurious symmetry-breaking terms can be removed in the renormalization process, as can also be observed in lattice regularization, which causes spurious Lorentz symmetry breaking (see \cite{Schierholz:1983yn,Caracciolo} and references therein for a discussion). However, it is necessary to identify whether the symmetry-breaking term is physical or not. For instance, gauge symmetry is preferred instead of the chiral one in the chiral anomaly computation. But this choice is made according to experiments. On the other hand, since there is no evidence of SUSY yet, symmetry breaking can not be decided by experiments, and a regularization scheme that does not break SUSY is preferred. 

In this work, we study the breaking of the scale symmetry at the one-loop order in supersymmetric quantum Electrodynamics (SQED). In particular, we consider $\mathcal{N}=1$ SQED, with one matter flavor, represented by two oppositely charged chiral multiplets, in the component formalism written in terms of Weyl spinors. 
We note that, despite the superfield formalism being a very powerful tool for studying quantum effects in SUSY quantum field theories, it is much more convenient to study conservation laws within the component formalism framework, which we will employ in this paper.

The paper is organized as follows. In the next section, we present an overview of the scale symmetry of this model and its possible anomalous amplitudes. In section \ref{s3}, we present a summary of the regularization scheme employed. In section \ref{s4}, we perform the one-loop calculation of the energy-momentum trace in the massless limit. We also discuss the scale anomaly for the case of different
masses and considering the low- and high-energy limits in section \ref{s5}. Our conclusions are presented in section \ref{s6}.

\section{Scale symmetry and dilatation current in SQED}
\label{s2}

As it is well known, a theory displays scale invariance if it stays unchanged under the following symmetry transformations,

\be
x'= e^{\epsilon} x
\label{eq0}
\ee

and

\be
\Phi'(x')= e^{-\epsilon d_{\Phi}} \Phi(x),
\label{eq1}
\ee
where $\epsilon$ is a scale parameter, $\Phi$ is a generic field and $d_{\Phi}$ is its scale dimension.

As an example, the kinetic terms of the electromagnetic or spinor fields, $-\frac{1}{4}F^{\mu\nu}F_{\mu\nu}$ 
and $\bar{\psi}\gamma^{\mu}\partial_{\mu}\psi$, respectively, are scale invariant because they possess scale 
dimension equal to 4. In this way, when the volume element $d^4x$ changes under conformal 
transformations, it compensates the change in the fields, making the action scale invariant. On the 
other hand, mass terms or interaction terms with dimensionful coupling constants break scale symmetry at the 
classical level because they do not transform according to the scale transformations presented in the equations above. It is important not to confuse ordinary dimension with scale dimension. For instance, although masses have ordinary dimension equal to $1$, their scale dimension is zero.

The Noether current related to the transformations given by Eqs. (\ref{eq0}) and (\ref{eq1}) is the dilatation current $\mathcal{J}^{\mu}=\Pi^{\mu}_id_{\Phi_i}\Phi_i+x_{\nu}\Theta^{\mu\nu}$, where $\Pi^{\mu}_i=\frac{\partial \mathcal{L}}{\partial (\partial_{\mu}\Phi_i)}$ is the canonical conjugate momentum of the $i-th$ field and $\Theta^{\mu\nu}$ is the standard energy momentum tensor given by 
$\Theta^{\mu\nu}=\Pi^{\mu}_i\partial^{\nu}\Phi_i-g_{\mu\nu}\mathcal{L}$. It is not guaranteed that this 
energy-momentum tensor will be symmetric. Therefore, improvement terms are sometimes required, as presented
in this work in Appendix A, where the Belinfante tensor is built for QED and SQED.

Besides the ordinary dimensionful terms with no scale dimension, like mass terms in the Lagrangian, the renormalization procedure also breaks scale symmetry because it introduces an RG scale dependence in the couplings. This effect is known as the anomalous breaking of the scale symmetry. This is physically expected since the same theory considered at a larger scale is characterized by a different value of the renormalized coupling constant.

We consider the SQED in order to study the breaking of the dilatation symmetry through an explicit calculation of a divergence of the dilatation current, or the trace of the energy-momentum tensor. This theory is described by the following  Lagrangian written in terms of Weyl spinors:

\begin{align}
&\mathcal{L}_{SQED} =i\bar{\lambda}\bar{\sigma}^{\mu}\partial_{\mu}\lambda-\frac{1}{4}F^{\mu\nu}F_{\mu\nu}+(D^{\mu}\tilde{\phi}_e)^{\dagger}(D_{\mu}\tilde{\phi}_e)+\nonumber\\
&+i\bar{\chi_{e}}\bar{\sigma}^{\mu}D_{\mu}\chi_{e}+(\sqrt{2}e\chi_{e}\cdot \lambda\tilde{\phi}_e+h.c.)+(D^{\mu}\tilde{\phi}_{\tilde{e}}^{\dagger})(D_{\mu}\tilde{\phi}_{\tilde{e}})+\nonumber\\
&+i\bar{\chi}_{\tilde{e}}\bar{\sigma}^{\mu}D_{\mu}\chi_{\tilde{e}}-(\sqrt{2}e\chi_{\tilde{e}}\cdot \lambda\tilde{\phi}_{\tilde{e}}+h.c.)
-m\chi_{e}\cdot \chi_{\tilde{e}}-\nonumber\\
&-m\bar{\chi}_{e}\cdot \bar{\chi}_{\tilde{e}}-m^2|\tilde{\phi}_{\tilde{e}}|^2-m^2|\tilde{\phi}_e|^2-\frac{e^2}{2}(|\tilde{\phi}_{\tilde{e}}|^2-|\tilde{\phi}_e|^2)^2,
\label{EQ1}
\end{align}
where $D_{\mu}\equiv \partial_{\mu}+ieA_{\mu}$ is the usual covariant derivative which couples the gauge field with matter. $\chi_{e}$,
$\chi_{\tilde{e}}$, $\tilde{\phi}_{e}$ and $\tilde{\phi}_{\tilde{e}}$ are spinor and scalar fields constituting the chiral multiplet, together with conjugated ones; $F^{\mu\nu}$ is the usual strength tensor of the photon field and $\lambda$ is the photino field. All the spinors are Weyl ones and left-chiral.

We expect scale symmetry to be broken at the classical level by all mass terms, {\it i. e.} the ones for the electron and the ones for the selectron. All the other fields in the Lagrangian have scale dimension equal to $4$, and therefore they are scale invariant.
   
As in the usual electrodynamics, it is possible to construct a symmetric energy-momentum tensor, known as the Belinfante tensor, whose trace leads to the classical breaking of the dilatation current conservation law. We explicitly show in Appendix A that the divergence of the dilatation current in the classical theory is given by
\be
\partial_{\mu}\mathcal{J}^{\mu}=T^{\mu}_{\ \ \mu}=m\chi_{\tilde{e}}\cdot\chi_e+m\bar{\chi}_{\tilde{e}}\cdot\bar{\chi}_e
+2m^2|\tilde{\phi}_{e}|^2+2m^2|\tilde{\phi}_{\tilde{e}}|^2,
\ee
where $T^{\mu}_{\ \ \mu}$ is the trace of the symmetric energy-momentum tensor.

The quantum corrections also break the dilatation current conservation. Applying the scale transformations of Eqs. (\ref{eq0}) and (\ref{eq1}), we find that the Ward identity of this dilatation current is given by
\begin{align}
&-i G_T(0,p_1,p_2,...,p_{n-1})=\Big(n(d_{\Phi}-4)+4-\nonumber\\
&-\sum_k^{n-1}p_k\frac{\partial}{\partial p_k}\Big)G(p_1,p_2,...,p_{n-1}),
\label{eqIW}
\end{align}
where $G_T(0,p_1,p_2,...,p_{n-1})$ and $G(p_1,p_2,...,p_{n-1})$ are $(n-1)$-point Green functions in the momentum space, representing themselves as Fourier transforms of the respective
coordinate space $n-$point Green functions, $G_T(y,x_1,x_2,...,x_{n})=\langle T^{\mu}_{\ \ \mu}(y)\Phi(x_1)\Phi(x_2)...\Phi(x_{n})\rangle$ and $G(x_1,x_2,...,x_{n})=\langle 
\Phi(x_1)\Phi(x_2)...\Phi(x_{n})\rangle$. Notice that the lower index $T$ is used to emphasize that one of the fields in the expectation value is the trace of the energy-momentum tensor. Besides, it is straightforward to show that the functional integral measure is invariant, $\mathcal{D}\Phi=\mathcal{D}\Phi'$, for a scale transformation within the functional integral. Since this is not true for quantum corrections, we are going to see in what follows that this equation is incomplete, {\it i.e.} there is an additional anomalous term in it, as it was first predicted by Fujikawa \cite{Fujikawa, Fujikawa2}.

For $n=2$ and $\Phi$ being the vector field, eq. (\ref{eqIW}) reads
\bq
&-i G^{\mu\nu}_T(0,p)=\left(n(d_{\Phi}-4)+4-p^{\lambda}\frac{\partial}{\partial p^{\lambda}}\right)G^{\mu\nu}(p), \nonumber\\
&-i G^{\mu\nu}_T(0,p)=\left(-2-p^{\lambda}\frac{\partial}{\partial p^{\lambda}}\right)G^{\mu\nu}(p).
\label{eqIWG}
\eq

It is also possible to rewrite (\ref{eqIWG}) in terms of the one-particle-irreducible (1PI) function $\Gamma^{\sigma\lambda}(p)$, related with the legged 2-point function
via the equation $G_{\mu\nu}(p)=D_{\mu\sigma}(p)\Gamma^{\sigma\lambda}(p)D_{\lambda\nu}(p)$, with $D_{\mu\nu}(p)=-\frac{ig_{\mu\nu}}{p^2}$ the photon propagator. The Ward
identity in eq. (\ref{eqIWG}) can be written as

\be
-i \Gamma^{\mu\nu}_T(0,p)=\left(2-p^{\lambda}\frac{\partial}{\partial p^{\lambda}}\right)\Gamma^{\mu\nu}(p).
\ee

\begin{figure}
\center
\subfigure[]{\includegraphics[trim=0mm 0mm 0mm 10mm,scale=0.45]{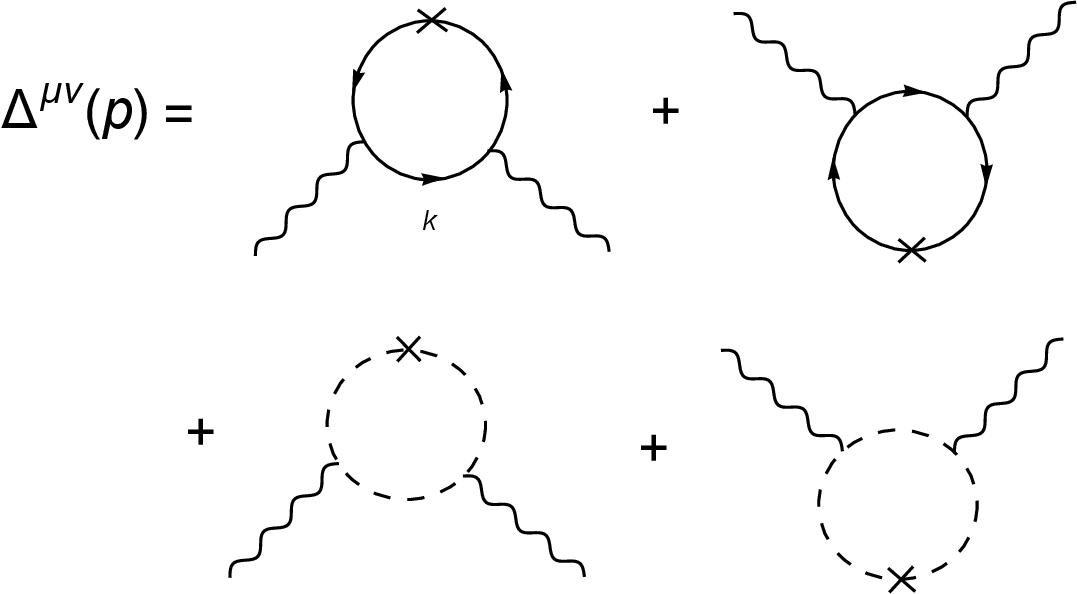}}
\subfigure[]{\includegraphics[trim=0mm 0mm 0mm 0mm,scale=0.45]{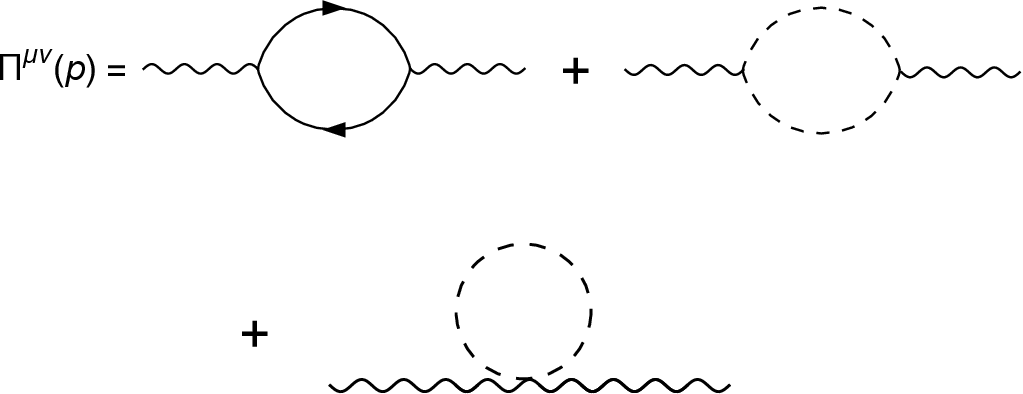}}
\caption{(a) One-loop 3-point Feynman diagram with the energy-momentum tensor as a vertex represented by a cross. (b) Vacuum polarization tensor.}
\label{fig1}
\end{figure}

Finally, $\Gamma^{\mu\nu}_T(0,p)$ corresponds to the 3-point Feynman diagram $\Delta^{\mu\nu}(0,p,-p)$ shown in Fig. \ref{fig1}, where two of the vertices are the usual ones
and the insertion is the trace of the energy-momentum tensor $T^{\mu}_{\ \ \mu}$,  $\Gamma^{\mu\nu}(p)$ corresponds to the 2-point Feynman diagram $\Pi^{\mu\nu}(p,-p)$, 
which is the vacuum polarization tensor. In terms of these diagrams the Ward identity can be written as:
\be
\Delta^{\mu\nu}=\left(2-p^{\lambda}\frac{\partial}{\partial p^{\lambda}}\right)\Pi^{\mu\nu}(p).
\label{eqIWD}
\ee

According to lagrangian (\ref{EQ1}), the solid and the dashed lines in the amplitudes of Fig. \ref{fig1} stand for the electron and the selectron propagator. The electron-photon interaction is $-ie\gamma_{\mu}$ while the selectron-photon interactions are $ie(p+p')_{\mu}$ if it comes from the 3-point Green function 
$\langle\tilde{\phi}_{e}^{\dagger}\tilde{\phi}_{e}A^{\mu}\rangle$ and
$2ie^2g^{\mu\nu}$ if it comes from the 4-point Green function $\langle\tilde{\phi}_{e}^{\dagger}\tilde{\phi}_{e}A^{\mu}A^{\nu}\rangle$.

For $n=2$ and $\Phi$ being the electron, photino and the selectron fields, the scale dimensions are
$d_{\chi}=3/2$, $d_{\lambda}=1$ and $d_{\phi}=1$, respectively. Thus, eq. (\ref{eqIW}) reads, for each case:
\be
\Delta_{\chi}=\left(1-p^{\lambda}\frac{\partial}{\partial p^{\lambda}}\right)\Sigma_{\chi}(p),
\label{eqIWD2}
\ee
\be
\Delta_{\lambda}=\left(1-p^{\lambda}\frac{\partial}{\partial p^{\lambda}}\right)\Sigma_{\lambda}(p),
\label{eqIWD3}
\ee
and
\be
\Delta_{\phi}=\left(2-p^{\lambda}\frac{\partial}{\partial p^{\lambda}}\right)D_{\phi}(p),
\label{eqIWD4}
\ee
where $\Sigma_{\chi}$, $\Sigma_{\lambda}$ and $D_{\phi}$ are the 1-loop self-energy of the electron, photino
and selectron, respectively. Unlike the photon case, none of the above Ward identities effectively contributes to the scale anomaly in the massless limit because the diagrams lead to terms proportional to the mass or $\slashed{p}$. The former gives no contribution to the Ward identity and the latter produces an equation of motion operator. In this way, mass-like terms such as $m\chi_e\cdot\chi_{\tilde{e}}$ or $m^2|\tilde{\phi}_e|^2$ contribute to the scale anomaly only classically because the naive Ward identities are not affected by the RG scale in the massless limit, as presented in table \ref{tab1}.

Since one-loop diagrams require a regularization and a renormalization procedure, a scale is introduced in this process and computing the diagrams of both sides of eq. (\ref{eqIWD}) yields the quantum breaking of scale symmetry expressed by the non-zero trace of the energy-momentum tensor.


\section{Implicit Regularization}
\label{s3}

We apply the implicit regularization (IReg) framework (for general information on it, see e.g. \cite{Orimar}-\cite{Luciana}) to treat the divergent integrals within our study. In this scheme, an implicit regularization parameter $\Lambda$ (not necessarily a hard cutoff), is assumed so that it makes sense to perform algebraic operations with the integrands. It is possible to use,
for example, the following identity to separate UV basic divergent integrals (BDI) from the finite part:
\begin{eqnarray}
\label{2.1}
&&\int_k \frac{1}{(k+p)^2-m^2}=\int_k\frac{1}{k^2-m^2}-\nonumber\\
&-&\int_k\frac{(p^2+2p\cdot k)}{(k^2-m^2)[(k+p)^2-m^2]},
\end{eqnarray}
where $\int_k\equiv\int^\Lambda\frac{d^4 k}{(2\pi)^4}$. These BDI's are defined as follows:
\begin{equation}
I^{\mu_1 \cdots \mu_{2n}}_{log}(m^2)\equiv \int_k \frac{k^{\mu_1}\cdots k^{\mu_{2n}}}{(k^2-m^2)^{2+n}},
\end{equation}
and
\begin{equation}
I^{\mu_1 \cdots \mu_{2n}}_{quad}(m^2)\equiv \int_k \frac{k^{\mu_1}\cdots k^{\mu_{2n}}}{(k^2-m^2)^{1+n}}.
\end{equation}

The basic divergences with Lorentz indices can be combined as differences between integrals with the same superficial degree 
of divergence, according to the equations below, which define surface terms \footnote{The Lorentz indices between brackets stand for 
permutations, i.e. $A^{\{\alpha_1\cdots\alpha_n\}}B^{\{\beta_1\cdots\beta_n\}}=A^{\alpha_1\cdots\alpha_{n}}B^{\beta_1\cdots\beta_n}$ 
+ sum over permutations between the two sets of indices $\alpha_1\cdots\alpha_{n}$ and $\beta_1\cdots\beta_n$.}:
\begin{align}
&\Upsilon^{\mu \nu}_{2w}= \eta^{\mu \nu}I_{2w}(m^2)-2(2-w)I^{\mu \nu}_{2w}(m^2) 
\equiv \upsilon_{2w}\eta^{\mu \nu},
\label{dif1}\\
\nonumber\\
&\Xi^{\mu \nu \alpha \beta}_{2w}=  \eta^{\{ \mu \nu} \eta^{ \alpha \beta \}}I_{2w}(m^2)
 -\nonumber\\&- 4(3-w)(2-w)I^{\mu \nu \alpha \beta }_{2w}(m^2) \equiv  \xi_{2w}\eta^{\{ \mu \nu} \eta^{ \alpha \beta \}},
\label{dif2}.
\end{align}

In the expressions above, $2w$ is the degree of divergence of the integrals, and,  for the sake of brevity, we substitute the subscripts 
$log$ and $quad$ by $0$ and $2$, respectively. Surface terms can be conveniently written as integrals of total 
derivatives, namely
\be
\upsilon_{2w}\eta^{\mu \nu}= \int_k\frac{\partial}{\partial k_{\nu}}\frac{k^{\mu}}{(k^2-m^2)^{2-w}}
\label{ts1}
\ee
and
\be
(\xi_{2w}-v_{2w})\eta^{\{ \mu \nu} \eta^{ \alpha \beta \}}= \int_k\frac{\partial}{\partial k_{\nu}}\frac{2(2-w)k^{\mu} k^{\alpha} k^{
\beta}}{(k^2-m^2)^{3-w}}.
\label{ts2}
\ee

Equation (\ref{2.1}) is applied recursively until the BDI's are separated from the finite parts. Also, we see that equations (\ref{dif1})-(\ref{dif2}) are undetermined because they represent themselves as differences between divergent quantities. Each regularization scheme gives a different value for these terms. However, as physics should not depend on the regularization, we leave these terms to be arbitrary until the end of the calculation, fixing them by symmetry constraints, like a Ward identity \cite{Jackiw2}.

The prescription above is not the usual regularization procedure and it is particularly useful in the computation of
anomalies, where we need to know if the quantum breaking of a symmetry really occurs or if it is a spurious symmetry breaking caused
by the regulator.

\section{One-loop breaking of the dilatation current}
\label{s4}

The amplitudes of the diagrams depicted at Fig. \ref{fig1} can be easily constructed using the Feynman rules. Their explicit form is given below:
\begin{eqnarray}
&&\Delta_{\mu\nu}=-2\int \frac{d^4k}{(2\pi )^4} Tr[e \gamma_{\mu}S(k-p)e \gamma_{\nu}S(k)m S(k)]+\nonumber\\&+&
8ie^2m^2\int \frac{d^4k}{(2\pi )^4}\frac{(2k-p)^{\nu}(2k-p)^{\mu}}{(k^2-m^2)^2[(k-p)^2-m^2]};
\nonumber\\
&&i\Pi^{\mu\nu}= \int \frac{d^4k}{(2\pi )^4} Tr[e \gamma_{\mu}S(k)e \gamma_{\nu} S(k+p)]-\nonumber\\
&-&2\int  \frac{d^4k}{(2\pi )^4} \frac{e^2(2k-p)^{\nu}(2k-p)^{\mu}}{(k^2-m^2)[(k-p)^2-m^2]}+\nonumber\\ &+&
2g^{\mu\nu}\int \frac{d^4k}{(2\pi )^4} \frac{2e^2}{k^2-m^2} ,
\label{Amps}
\end{eqnarray}
where $S(k)=i/(\slashed{k}-m)$ is the fermion propagator. There is a multiplicity factor $2$ in diagrams describing the selectron loop, corresponding to each superpartner associated with chiral electron states ($\chi_e$ and $\chi_{\tilde{e}}$). 

We list all finite and regularized divergent integrals in Appendix B. The results for the amplitudes in eq. (\ref{Amps}) are given by
\bq
&\Delta_{\mu\nu}=\frac{m^2 e^2}{2\pi^2}(g_{\mu\nu}Z_0(m^2)+\iota_0 g_{\mu\nu}p^2-2\iota_1(m^2) p_{\mu}p_{\nu}-\nonumber\\
&-16\pi^2 i\upsilon_0 g_{\mu\nu})-\frac{e^2m^2}{\pi^2}\frac{Z_0(m^2)}{p^2}(g^{\mu\nu}p^2-p^{\mu}p^{\nu})+\nonumber\\
&+4i e^2 m^2 g^{\mu\nu}(I_{log}(m^2)-\upsilon_0 ),
\label{eqdelta}
\eq
and
\bq
\centering
i\Pi^{\mu\nu}&=2(p^2g^{\mu\nu}-p^{\mu}p^{\nu})(I_{log}(m^2)-b Z_0(m^2))-\nonumber\\
&-2p^2g^{\mu\nu}\upsilon_0 ,
\label{eqpi}
\eq
where $Z_n(m^2)=\int_0^1 dx x^n\ln \left(\frac{\mathscr{D}(x)}{m^2}\right)$, $\iota_n(m^2)=\int_0^1 dx \frac{x^n(1-x)}{\mathscr{D}(x)}$ and $\mathscr{D}(x)=m^2-p^2 x(1-x)$.

We see in eq. (\ref{eqpi}) that the logarithmic surface term $\upsilon_0$ breaks gauge symmetry and should 
be set to zero, as we can easily see by requiring gauge symmetry, imposing the gauge Ward identity 
$p_{\mu}\Pi^{\mu\nu}=0$. After setting this surface term to vanish, the amplitude turns out to be transverse 
and gauge invariant. It is interesting to notice that unlike usual QED \cite{Bruno}, there is only one 
remaining surface term, which is not gauge invariant, and we can get rid of it by using the Ward identity. 

In eqs. (\ref{eqdelta}) and (\ref{eqpi}), the RG scale $\lambda$ is introduced with the use of the scale relation below:
\be
I_{log}(m^2)=I_{log}(\lambda^2)+b \ln \left(\frac{\lambda^2}{m^2}\right).
\ee

When the massless limit is taken, $\lambda$ is the only mass scale remaining in our results, and it can be
taken as the RG scale for massless theories.

The amplitude $\Delta_{\mu\nu}$ in eq. (\ref{eqdelta}) is zero in the massless limit for all terms. So, all the anomalous terms come from the amplitude $\Pi_{\mu\nu}$. Now, inserting Eqs. (\ref{eqdelta}) and
(\ref{eqpi}) in the naive Ward identity of the scale symmetry and taking the massless 
limit to restore the conservation of the dilatation current, we find the anomalous term
\begin{align}
&\Delta^{\mu\nu}_r-\Big(2-p^{\lambda}\frac{\partial}{\partial p^{\lambda}}\Big)\Pi^{\mu\nu}_r= \frac{e^2}{4\pi^2}(p^2g^{\mu\nu}-p^{\mu}p^{\nu}),
\label{result}
\end{align}
where the index $r$ in the tensors stands for the renormalized amplitudes. There is an additional gauge-invariant term if we compare Eq. (\ref{result}) with the result for the usual QED obtained in other regularization schemes \cite{Ellis, LoopReg, Porto}. This additional term corresponds to the contribution of the superpartners. We see that scale symmetry is broken by a gauge-invariant term with no ambiguity related to quadratic surface terms, as expected, since our Susy theory is free of quadratic divergences.
However, this condition is obtained only for $m=M$, as we are going to discuss in the next section.

Table \ref{tab1} summarizes how scale symmetry is broken for both classical and quantum levels. Since
the Ward identity of the vector field is the only one affected by the RG scale in the massless limit, the quantum breaking occurs only for this term. Actually, in the massive field limit, the classical breaking
terms also acquire quantum corrections because of the self-energy diagrams, which give corrections to the
masses.

\begin{table}
\centering
 \begin{tabular}{c c c c}
  \hline
  Lagrangian & Type of& Scale \\
	term &  breaking& dimension \\
  \hline
  $m\chi_{\tilde{e}}\cdot\chi_e$ & Classical& 3 \\
  \hline
   $m\bar{\chi}_{\tilde{e}}\cdot\bar{\chi}_e$& Classical&3 \\
  \hline
  $m^2|\tilde{\phi}_{e}|^2$& Classical&2 \\
  \hline
  $m^2|\tilde{\phi}_{\tilde{e}}|^2$& Classical&2\\
  \hline
  $e^2 F^{\mu\nu}F_{\mu\nu}$& Quantum&4 \\
  \hline 
 \end{tabular}
 \caption{Summary of the classical and quantum breaking of scale symmetry. If the theory is not massless,
 there is an additional correction to the masses because of the loop corrections and the mass terms also
 break scale symmetry in the quantum regime.}
\label{tab1} 
\end{table}

Eq. (\ref{result}) can be written in terms of fields, and it leads to the following trace of the energy-momentum tensor together with the classical breaking:
\begin{align}
T^{\mu}_{\ \ \mu}|_{classical}&=m\chi_{\tilde{e}}\cdot\chi_e+m\bar{\chi}_{\tilde{e}}\cdot\bar{\chi}_e+2m^2|\tilde{\phi}_{e}|^2+\nonumber\\
&+2m^2|\tilde{\phi}_{\tilde{e}}|^2,\nonumber\\
T^{\mu}_{\ \ \mu}|_{quantum}&=\frac{e^2}{16\pi^2}F^{\mu\nu}F_{\mu\nu},
\end{align} 
where we can notice that the coefficient of the quantum breaking $\frac{e^2}{16\pi^2}$ is proportional to the beta function as in the usual QED \cite{Duncan}: $\frac{\beta_{SQED}}{2e}$. Unlike the classical symmetry breaking terms, this one appears because of the renormalization process, which unavoidably introduces a scale $\lambda$ in the theory. The coefficient of the operator in the energy-momentum trace has the same form as in other theories and frameworks like usual QED \cite{Ellis, LoopReg, Porto} and the Lorentz-violating QED \cite{Porto}, where we find $\frac{\beta_{g}}{2g}$, with $g$ the corresponding coupling for the specific context.

In this way of computing the $\beta$-function, there is no ambiguity associated with surface terms 
\cite{Adriano}. At one-loop, this agrees with the NSVZ result expected for the beta function considering
$\mathcal{N}=1$, $N_f=1$ and $\gamma=0$ \cite{NSVZbf}. The running of the coupling following from this beta function gives

\be
e^2(\lambda)=\frac{e^2(\lambda_0)}{(1-\frac{e^2(\lambda_0)}{4\pi^2} \ln\frac{\lambda}{\lambda_0} )}.
\ee

Thus, for positive $\beta_{SQED}$, the perturbative running coupling increases toward the ultraviolet and reaches a Landau pole at a finite scale $\lambda_L$. Consequently, the limit $\lambda\rightarrow \infty$ cannot be taken within the perturbative SQED description. The quantum contribution to the trace vanishes instead at an RG fixed point satisfying $\beta_{SQED}=0$.

\section{The trace anomaly with softly broken Susy} 
\label{s5}

The trace anomaly in QED is reproduced for both high- and low-energy limits besides the usual massless limit. We present the result for the diagram $\Delta^{\mu\nu}$ and the operator $\left(2-p^{\lambda}\frac{\partial}{\partial p^{\lambda}}\right)\equiv \mathcal{O}$ acting on the vacuum polarization tensor together with the trace anomaly coefficient for each one of these cases in Table \ref{tab2}. This table is obtained from the explicit results below for QED only:

\begin{table}
\centering
 \begin{tabular}{c c c c}
  \hline
  Tensor & $m\rightarrow 0$& $m^2>>p^2$& $m^2<<p^2$\\
  \hline
  $\mathcal{O}\Pi^{\mu\nu}$ & $-\frac{e^2}{6\pi^2}\mathcal{P}^{\mu\nu}$& 0& $-\frac{e^2}{6\pi^2}\mathcal{P}^{\mu\nu}$  \\
  \hline
  $\Delta^{\mu\nu}$ & 0& $\frac{e^2}{6\pi^2}\mathcal{P}^{\mu\nu}$ & 0  \\
  \hline
  Coeff. in $T^{\mu}_{\ \ \mu}$& $\frac{\beta_{QED}}{2e}$& $\frac{\beta_{QED}}{2e}$& $\frac{\beta_{QED}}{2e}$\\
  \hline 
 \end{tabular}
 \caption{Summary of the quantum breaking of scale symmetry in the massless, low and high energy limits, where $\mathcal{P}^{\mu\nu}=(p^2g^{\mu\nu}-p^{\mu}p^{\nu})$, for QED.}
\label{tab2} 
\end{table}

\bq
&\Delta^{\mu\nu}_r=\frac{m^2 e^2}{2\pi^2}(g^{\mu\nu}Z_0+\iota_0 g^{\mu\nu}p^2-2\iota_1 p^{\mu}p^{\nu}),
\label{eqdelta3}
\eq
and
\bq
\centering
\mathcal{O}\Pi^{\mu\nu}_r&=\frac{e^2}{6\pi^2}\mathcal{P}^{\mu\nu}\left( (2m^2+p^2)\iota_1(m^2) +\frac{2m^2}{p^2}Z_0(m^2)\right)+\nonumber\\
&+ 8i\upsilon_2(m^2)g^{\mu\nu},
\label{eqpi2}
\eq
where $\mathcal{P}^{\mu\nu}=(p^2\eta^{\mu\nu}-p^{\mu}p^{\nu})$, $Z_n=\int_0^1 dx x^n\ln \left(\frac{\D(x)}{m^2}\right)$, $\iota_n=\int_0^1 dx \frac{x^n(1-x)}{\D(x)}$ and $\D(x)=m^2-p^2 x(1-x)$.
We can again use the gauge Ward identities to get $\upsilon_2(m^2)=0$.

The contribution from $\mathcal{O}\Pi^{\mu\nu}$ in eq. (\ref{eqpi}) disappears for a heavy fermion, and this is expected because of the Appelquist-Carazzone theorem \cite{ACT}, {\it i. e.} if the particle is much heavier than the energy scale, its effects are irrelevant (proportional to $p^2/m^2$). However, the insertion $m\bar{\psi}\psi$ transforms the mass operator into a local term that appears when the heavy fermion degree of freedom is integrated out of the theory in the limit $p^2/m^2\rightarrow 0$. So, the 
anomaly comes from eq. (\ref{eqdelta3}) in this limit. In the high energy limit ($p^2>>m^2$), the
situation is similar for the massless case except that we neglect terms proportional to $m^2/p^2$. Therefore, the result of the scale anomaly is the same for the massless, low- and high-energy limits.

Now, we consider soft symmetry breaking of SUSY and therefore different masses for the selectron and the 
electron. As in the usual QED, we are going to evaluate the low- and the high-energy limits for SQED. The full results for $\mathcal{O}\Pi^{\mu\nu}$ and $\Delta^{\mu\nu}$ considering $m\neq M$ are respectively:
\begin{eqnarray}
&&\mathcal{O}\Pi^{\mu\nu}_r=\frac{e^2}{12\pi^2}(p^2g^{\mu\nu}-p^{\mu}p^{\nu})\Big[ p^2(2\iota_1(m^2)+\iota_1(M^2))+\nonumber\\
&+&\frac{4m^2}{p^2}(p^2\iota_1(m^2)+Z_0(m^2))- \frac{4M^2}{p^2}(p^2\iota_1(M^2)+Z_0(M^2))\Big] \nonumber\\
&+&8g^{\mu\nu}(\upsilon_2(m^2)-\upsilon_2(M^2))
\label{eqDPi}
\end{eqnarray}
and
\bq
&\Delta^{\mu\nu}_r=\frac{m^2 e^2}{2\pi^2}(g^{\mu\nu}Z_0(m^2)+\iota_0(m^2) g^{\mu\nu}p^2-2\iota_1(m^2) p^{\mu}p^{\nu})-\nonumber\\
&-\frac{e^2M^2}{\pi^2}\frac{Z_0(M^2)}{p^2}(g^{\mu\nu}p^2-p^{\mu}p^{\nu}),
\label{eqdelta2}
\eq
where $m$ and $M$ are the electron and the selectron masses, respectively.

As expected, the quadratic surface term does not vanish this time since the masses are different. Nevertheless, this term breaks gauge symmetry. It can be set to zero by requiring gauge invariance with the use of the Ward identity $p_{\mu}\Pi^{\mu\nu}=0$. In this way, explicit soft SUSY breaking requires gauge
symmetry to get rid of quadratic surface terms. This requirement implies in $\upsilon_2(m^2)=\upsilon_2(M^2)$ or $\upsilon_2(m^2)=0$ and $\upsilon_2(M^2)=0$.

\begin{table}
\centering
 \begin{tabular}{c c c c}
  \hline
  Tensor & $m= M$ & $M^2<<p^2$& $M^2>>p^2$\\
         & and $m\rightarrow 0$& and $m^2<<p^2$& and $m\rightarrow 0$\\
  \hline
  $\mathcal{O}\Pi^{\mu\nu}_{\tilde{e}}$ & $-\frac{e^2}{12\pi^2}\mathcal{P}^{\mu\nu}$& $-\frac{e^2}{12\pi^2}\mathcal{P}^{\mu\nu}$ & 0 \\
  \hline
  $\Delta^{\mu\nu}_{\tilde{e}}$ & 0& 0 & $\frac{e^2}{6\pi^2}\mathcal{P}^{\mu\nu}$  \\
  \hline
  $\mathcal{O}\Pi^{\mu\nu}_{e}$ & $-\frac{e^2}{6\pi^2}\mathcal{P}^{\mu\nu}$& $-\frac{e^2}{6\pi^2}\mathcal{P}^{\mu\nu}$ & $-\frac{e^2}{6\pi^2}\mathcal{P}^{\mu\nu}$  \\
  \hline
  $\Delta^{\mu\nu}_{e}$ & 0& 0 & 0  \\
  \hline
  Coefficient in $T^{\mu}_{\ \ \mu}$& $\frac{\beta_{SQED}}{2e}$& $\frac{\beta_{SQED}}{2e}$& $\frac{\beta_{QED}}{2e}$\\
  \hline 
 \end{tabular}
 \caption{Summary of the quantum breaking of scale symmetry in the massless, low and high energy limits, where $\mathcal{P}^{\mu\nu}=(p^2g^{\mu\nu}-p^{\mu}p^{\nu})$, for SQED. In the last column, the local contribution $\Delta_{\tilde{e}}^{\mu\nu}$ is interpreted as a threshold matching contribution and is not added to the infrared dynamical QED contribution.}
\label{tab3} 
\end{table}

We see again that heavy degrees of freedom are decoupled so that $\mathcal{O}\Pi^{\mu\nu}_e$ leads to the usual result expected for QED. Decoupling the selectron produces the expected result for the
anomaly without the need of considering diagram $\Delta^{\mu\nu}_e$, as before. The diagram $\Delta^{\mu\nu}_{\tilde{e}}$ by its turn also leads to the QED anomaly in the limit $M^2>>p^2$, while
$\mathcal{O}\Pi^{\mu\nu}_{\tilde{e}}\rightarrow 0$. Unlike the previous result for QED, the Appelquist-Carazzone theorem \cite{ACT} shows that the effect of the heavy particle ($M^2>>p^2$) is absorbed in parameters of the effective field theory, the ones related to the electron.
In this way, both Ward identities for the electron and the selectron, {\it i. e.} $\Delta^{\mu\nu}_e-\mathcal{O}\Pi^{\mu\nu}_e$ and $\Delta^{\mu\nu}_{\tilde{e}}-\mathcal{O}\Pi^{\mu\nu}_{\tilde{e}}$, 
reproduce the same matching relation in the heavy-selectron limit. In this limit, the electron contribution remains dynamical and reproduces the QED anomaly as before, whereas the selectron contribution is absorbed into local matching coefficients of the low-energy effective theory. Therefore, the two Ward identities should not be interpreted as independent contributions to be summed.

\section{Conclusions and perspectives} 
\label{s6}

We study the impacts of quantum corrections on a scale symmetry within the SQED formulated within the component framework. The results show that operators of the SQED Lagrangian break scale symmetry at the classical level if they do not possess a scale dimension equal to 4. As expected, these terms are mass terms only because all the coupling constants are dimensionless. Also, since the renormalization process introduces a scale, we also have in the massless limit the quantum breaking of the energy-momentum tensor trace due to dimensionless coefficients when quantum corrections are taken into account. As in the usual QED, this scale-breaking term is gauge invariant although it has a different coefficient.

From an experimental point of view, the scale anomaly is directly related to the hadronic ratio
$R_{had}=\frac{\sigma(e^+e^-\rightarrow \gamma\rightarrow hadrons)}{\sigma(e^+e^-\rightarrow \gamma\rightarrow \mu^-\mu^+)}$ \cite{Ellis} in the high-energy limit. In our present situation, the equivalent observable is $R=\frac{\sigma(e^+e^-\rightarrow \gamma\rightarrow \tilde{l}^+\tilde{l}^-,...)}{\sigma(e^+e^-\rightarrow \gamma\rightarrow \mu^-\mu^+)}$. Thus, available measurements of the $R$ ratio offer the possibility of indirect testing SUSY since the coefficient of the scale symmetry breaking term is different from the one provided by the usual QED. Nevertheless, this effect is suppressed because of the heaviness of the superpartners. The ALEPH collaboration experiment excludes light gluinos \cite{ALEPH} or, more specifically, in the case of this work, there is exclusion of light selectrons \cite{ALEPH2}. Recently, the ATLAS collaboration probes sleptons' masses up to the 250-350 GeV range in compressed scenarios, while stronger limits are obtained for favorable mass spectra \cite{ATLAS}, which means that $M^2>>p^2$ and the result for usual QED in the trace anomaly is expected because $R\rightarrow R_{QED}$ in available measurements from electron colliders. The limit where we would expect 
$T^{\mu}_{\ \ \mu}=\frac{\beta_{SQED}}{2e}\rightarrow \frac{R_{SQED}}{R_{QED}}=\frac{3}{2}$ requires 
$s=p^2>>M^2$, like $\sqrt{s}\sim 2\ TeV$ for $M\sim\ 300\ GeV$.

A possible generalization of our study could consist in applying this methodology to the LV SQED, originally formulated in \cite{Lehum}, which provides different operators with scale-dimensionful coefficients, not necessarily mass terms because several LV coefficients
have ordinary dimension and no scale dimension. We will perform this study in one of our forthcoming papers.

\section*{Acknowledgements}
The work of A. Yu.\ P. has been partially supported by the CNPq project No. 303777/2023-0.

\appendix

\section{Classical breaking of scale symmetry}
\label{A1}

In this Appendix, we show how the classical terms breaking the scale symmetry arise from the SQED Lagrangian. The case for the scalar field $\tilde{\phi}_e$ corresponding to the selectron is trivial. Considering a Lagrangian for a complex scalar field, $\mathcal{L}=\partial_{\mu}\phi^{\dagger}\partial^{\mu}\phi-m^2\phi^{\dagger}\phi$, we have
$\Pi^{\mu}=\frac{\partial \mathcal{L}}{\partial(\partial_{\mu}\phi)}=\partial^{\mu}\phi^{\dagger}$ and $\Theta^{\mu\nu}=\partial^{\mu}\phi^{\dagger}\partial^{\nu}\phi+\partial^{\nu}\phi^{\dagger}\partial^{\mu}\phi-g^{\mu\nu}\mathcal{L}$. Consequently, one finds $\Theta^{\mu}_{\ \ \mu}=2m^2|\phi|^2$ after the use of the equations of motion and the dropping of total derivatives.

For the Dirac and electromagnetic fields, the canonical energy-momentum tensor is
\begin{align}
&\Theta^{\mu\nu}=\frac{\partial \mathcal{L}}{\partial(\partial_{\mu}\psi)}\partial^{\nu}\psi+\frac{\partial \mathcal
{L}}{\partial(\partial_{\mu}\bar{\psi})}\partial^{\nu}\bar{\psi}+\frac{\partial \mathcal{L}}{\partial(\partial_{\mu
}A_{\alpha})}\partial^{\nu}A_{\alpha}- \nonumber\\
&-g^{\mu\nu}\mathcal{L}=i\bar{\psi}\gamma^{\mu}\partial^{\nu}\psi+F^{\lambda\mu}\partial^{\nu}A_{\lambda}-g^{\mu\nu}\mathcal{L}.
\label{EMT}
\end{align}

Although this canonical tensor is not symmetric, it can be symmetrized. The symmetric version is known as the Belinfante tensor:
\be
T^{\mu\nu}= \Theta^{\mu\nu}+\partial_{\rho}f^{\mu\nu\rho}_A+\partial_{\rho}f^{\mu\nu\rho}_{\psi},
\ee
where $f^{\mu\nu\rho}_{\Phi}=\frac{1}{2}(\Pi^{\rho}S^{\mu\nu}\Phi-\Pi^{\mu}S^{\rho\nu}\Phi+\Pi^{\nu}S^{\mu\rho}\Phi)$ and $S^{\mu\nu}$ is the spin matrix associated with 
the field $\Phi$.

Computing $\partial_{\rho}f^{\mu\nu\rho}$ for the spinor and the vector fields, we find
\begin{align}
&\partial_{\rho}f^{\mu\nu\rho}_{A}=e\bar{\psi}\gamma^{\mu}\psi A^{\nu}+F^{\mu\rho}\partial_{\rho}A^{\nu}, \nonumber\\
&\partial_{\rho}f^{\mu\nu\rho}_{\psi}=\frac{i}{4}(\bar{\psi}\gamma^{\nu}\partial^{\mu}\psi-(\partial^{\mu}\bar{\psi})\gamma^{\nu}\psi-3\bar{\psi}\gamma^{\mu}\partial^{\nu}\psi -\nonumber\\
&-(\partial^{\nu}\bar{\psi})\gamma^{\mu}\psi)+\frac{1}{2}e\bar{\psi}(\gamma^{\nu}A^{\mu}-\gamma^{\mu}A^{\nu})\psi.
\label{eqf}
\end{align}

Eqs. (\ref{eqf}) leads to the symmetric form of the energy-momentum tensor
\begin{align}
&T^{\mu\nu}=\frac{i}{4}[\bar{\psi}\gamma^{\mu}D^{\nu}\psi-\nonumber\\
&-(\overline{D^{\mu}\psi})\gamma^{\nu}\psi+\bar{\psi}\gamma^{\nu}D^{\mu}\psi-(\overline{D^{\nu}\psi})\gamma^{\mu}\psi ] +\nonumber\\ 
&+F^{\lambda\mu}F_{\lambda}^{\nu}-g^{\mu\nu}\mathcal{L},
\end{align}
whose trace is $T^{\mu\nu}=m\bar{\psi}\psi$.

In SQED, the energy-momentum tensor can be obtained straightforwardly in the same way for Weyl fermions. Considering the fermion sector in the Lagrangian of eq. (\ref{EQ1}), the symmetric energy-momentum tensor is given by
\begin{align}
&T^{\mu\nu}=\frac{i}{4}[\bar{\chi}_e\bar{\sigma}^{\mu}D^{\nu}\chi_e-(D^{\mu}\overline{\chi}_e)
\bar{\sigma}^{\nu}\chi_e+\bar{\chi}_e\bar{\sigma}^{\nu}D^{\mu}\chi_e- \nonumber\\ 
&-(D^{\nu}\overline{\chi}_e)\bar{\sigma}^{\mu}\chi_e]+(e\rightarrow \bar{e})-g^{\mu\nu}\mathcal{L}_f,
\end{align}

whose trace is
\be
T^{\mu}_{\ \ \mu}=\frac{i}{2}\bar{\chi}_e\bar{\sigma}^{\mu} \overleftrightarrow{D}_{\mu}\chi_e+
\frac{i}{2}\bar{\chi}_{\tilde{e}}\bar{\sigma}^{\mu} \overleftrightarrow{D}_{\mu}\chi_{\tilde{e}}-4\mathcal{L}_f
\label{traco}
\ee

Using the Lagrangian given by Eq.(\ref{EQ1}) and the equations of motion $i\bar{\sigma}^{\mu}D_{\mu}\chi_e=m\bar{\chi}_{\tilde{e}}$, $i\bar{\sigma}^{\mu}D_{\mu}\chi_{\tilde{e}}=m\bar{\chi}_e$, $iD_{\mu}\bar{\chi}_e\bar{\sigma}^{\mu}=-m\chi_{\tilde{e}}$ and $iD_{\mu}\bar{\chi}_{\tilde{e}}\bar{\sigma}^{\mu}=-m\chi_e$, we find out the classical divergence of the dilatation current for the fermion sector:
\be
T^{\mu}_{\ \ \mu}=m(\chi_{e}\cdot \chi_{\tilde{e}}+\bar{\chi}_{e}\cdot \bar{\chi}_{\tilde{e}}).
\ee

\section{Regularized integrals}
\label{A2}

The results of all finite and regularized divergent integrals for $\Delta^{\mu\nu}$ and $\Pi^{\mu\nu}$ amplitudes of section \ref{s4} are listed below:

\be
\int_k \frac{1}{(k^2-m^2)[(k+p)^2-m^2]^2}=-b \iota_0(m^2) ,
\ee
\be
\int_k \frac{k^{\mu}}{(k^2-m^2)^2[(k+p)^2-m^2]}=b p^{\mu} \iota_1(m^2) ,
\ee
\be
\int_k \frac{1}{[(k^2-m^2)}=I_{quad}(m^2), 
\ee
\be
\int_k \frac{1}{[(k+p)^2-m^2]}=I_{quad}(m^2)-p^2\upsilon_0(m^2), 
\ee
\begin{align}
&\int_k \frac{k^{\mu}k^{\nu}}{(k^2-m^2)^2[(k+p)^2-m^2]}=\nonumber\\
&=\frac{1}{4}g^{\mu\nu}(I_{log}(m^2)-\upsilon_0(m^2))-\nonumber\\
&-\frac{1}{2}b g^{\mu\nu}(Z_0(m^2)-Z_1(m^2))-b p^{\mu}p^{\nu}\iota_2(m^2) ,
\end{align}
\begin{align}
&\int_k \frac{k^{\mu}}{(k^2-m^2)[(k+p)^2-m^2]}=\nonumber\\
&=-\frac{1}{2}p^{\mu}(I_{log}(m^2)-\upsilon_0(m^2) -b Z_0(m^2)),
\end{align}
\begin{align}
\centering
&\int_k \frac{1}{(k^2-m^2)[(k+p)^2-m^2]}=\nonumber\\
&=I_{log}(m^2) -b Z_0(m^2),
\end{align}
\begin{align}
\centering
&\int_k \frac{k^{\mu}k^{\nu}}{(k^2-m^2)[(k+p)^2-m^2]}=\\
&=\frac{1}{2}g^{\mu\nu}(I_{quad}(m^2)-\upsilon_2(m^2) )+\nonumber\\
&+\left(\frac{1}{3}p^{\mu}p^{\nu}-\frac{1}{12}p^{2}\eta^{\mu \nu} \right)( I_{log}(m^2) -b Z_0(m^2))- \nonumber\\
&-\frac{1}{6}\xi_0(m^2)(p^2g^{\mu\nu}+2p^{\mu}p^{\nu}) +\frac{1}{4}g^{\mu\nu}p^2 \upsilon_0(m^2)-\nonumber\\
&-\frac{1}{3}b(p^2g^{\mu\nu}-p^{\mu}p^{\nu})\left(\frac{m^2}{p^2}Z_0(m^2) +\frac{1}{6} \right), 
\end{align}
where we use the notations: $Z_n(m^2)=\int_0^1 dx x^n \ln \frac{\mathscr{D}}{m^2}$, $\iota_n(m^2)=\int_0^1 dx \frac{x^n(1-x)}{\mathscr{D}}$, $b\equiv \frac{i}{(4\pi)^2}$ and $\mathscr{D}=\mathscr{D}(x)=m^2-p^2 x(1-x)$. Also, the following relation involving integrals in the Feynman parameters can be used to reduce finite integrals:
\begin{eqnarray}
&&Z_k(m^2)=\frac{k}{k+1} Z_{k-1}(m^2)
-\frac{k-1}{k+1}\frac{m^2}{p^2}Z_{k-2}(m^2)\nonumber\\
&-\frac{k-1}{k(k+1)^2},
\end{eqnarray}
where $k>0$.






\end{document}